\documentclass[twocolumn]{article}

\usepackage{framed}
\usepackage{hyperref}

\newcommand{\comment}[1]{}

\def\denseformat{
\setlength{\textheight}{9.2in}
\setlength{\textwidth}{7.1in}
\setlength{\evensidemargin}{-0.2in}
\setlength{\oddsidemargin}{-0.2in}
\setlength{\headsep}{10pt}
\setlength{\topmargin}{-0.3in}
\setlength{\columnsep}{0.375in}
\setlength{\itemsep}{0pt}
\renewcommand{\baselinestretch}{0.99}
}

\denseformat

\begin{document}
\bibliographystyle{plain}

\title{{\sc DiffSum} -- A Simple Post-Election Risk-Limiting Audit}
\author{Ronald L. Rivest\\
  MIT CSAIL, {\texttt{rivest@mit.edu}}}
\date{\today}
\maketitle

\thispagestyle{empty}

We present {\sc DiffSum}, a simple risk-limiting post-election ballot-polling audit.
See~\cite{NBHC07,LindemanStYa12,LindemanSt12}
for background.

You wish to check that candidate A really won a plurality election against candidate B.
You may sample the $n$ cast paper ballots without replacement.

\begin{leftbar}
\noindent {\bf Procedure {\sc DiffSum:}}
\begin{enumerate}

\item {\bf [Choose $c$]} Let $d$ be the number of decimal digits in $n$, and
  choose $c = d+\delta$ where $\delta$ controls the error rate (the chance of the audit accepting
  an incorrect outcome):
  \begin{tabular} {c|c|c|c|c|c}
      $\delta$          &  0   &  1   &  2   &  3   & 4 \\ \hline
      \hbox{max error rate} & 22\% & 15\% & 10\% &  6\% & 4\% 
    \end{tabular}

  \item {\bf [Begin]} Draw an initial sample of $24$ ballots.

  \item {\bf [Tally]} Determine the number $a$ of votes for $A$ in your sample, and the number $b$ of
    votes for $B$.

  \item {\bf [Stop?]} Stop the audit (accept A as winner) if $a>b$ and 
    \begin{equation}
         (a-b)^2 > c \cdot (a+b)\ .
    \label{eq:diffsum1}
    \end{equation}

  \item {\bf [Continue?]} 
    If $a+b=n$, stop (you have just completed a full recount).
    Otherwise, enlarge your random sample and return to step~3.
\end{enumerate}
\vspace{-0.15in}
\end{leftbar}

\noindent{\bf Remarks:} 
The initial size 24 of the sample in step 2 is arbitrary.
In step 5 the increase in sample size is also arbitrary; it could be by a
single ballot.

The name ``{\sc DiffSum}'' was chosen because
(\ref{eq:diffsum1}) says
\begin{equation}
     (\hbox{difference})^2 > c \cdot (\hbox{sum})\ .
  \end{equation}

\noindent{\bf Efficiency:} Let $m$ be the true margin (the
fraction of votes cast for A minus the fraction cast for B).
In a sample of size $s = a+b$,
the expected value of $a-b$ is $sm$.  Thus, {\sc DiffSum} is expected to
stop when
$ 
       (sm)^2 > cs
$
or
\begin{equation}
        s > c / m^2
\label{eq:s-bound}
\end{equation}

{\sc DiffSum} is approximately as efficient as {\sc Bravo}---compare
(\ref{eq:s-bound}) with the estimate $2 \ln(1/\alpha) / m^2$ for {\sc
  Bravo}~\cite{LindemanStYa12} (here $\alpha$ is the risk limit).
Moreover, {\sc DiffSum} does not need an initial estimate of the vote
shares, and {\sc Bravo} is inefficient when this estimate is
inaccurate.

\noindent{\bf Error rate:}
The error rate bounds given in Step 1 are based on
extensive simulations for 
$\delta=0$ to $4$, 
$d=3$ to $7$,
$n=10^d$, and $c=d+\delta$. 
We measured the error rate over 10,000 simulated elections
in each case.
Each simulation estimated the error rate when the election was
a tie, a worst-case scenario;
with more realistic margins the error
rate drops dramatically, so that in practice
even $c=d$ should give very reliable audits.

\smallskip
\noindent{\bf Example:} 
An election with $n=50,000$ votes can be audited using $c=7$ for 
a risk limit
of $\alpha=10\%$.
For $m=0.20$, {\sc DiffSum} examines about 175 ballots (estimated),
{\sc Bravo} (with $\alpha=0.10$) examines about 115 (estimated).
In simulations for this election,
{\sc DiffSum} with $c=7$
examines about 157 ballots on average, and has an error rate of less than 0.04\%;  
{\sc DiffSum} with $c=5$
examines about 112 ballots on average, and has an error rate of less than 0.2\%.
Bravo examines about 119 ballots on average, and has an error rate of approximately 2.5\%.

\smallskip
\noindent{\bf Extension:}
In practice, one should cease random sampling once a significant
number (say 4\%) of the ballots have been sampled, when switching over
to a full hand recount becomes more economical.

With more candidates, let {\sc DiffSum} check that the
sample winner beats the sample's strongest loser.

\smallskip
\noindent {\bf Conclusion:}
{\sc DiffSum} is exceptionally simple, and 
appears quite comparable to {\sc Bravo}
in terms of efficiency and error rate.  Further simulations and
analysis would be helpful.

\smallskip
\noindent {\bf Acknowledgment:} I thank Philip Stark for helpful comments.

\small
\vspace*{-0.2in}

\bibliography{audit}

\begin{thebibliography}{1}

\bibitem{LindemanSt12}
M.~Lindeman and P.~B. Stark.
\newblock A gentle introduction to risk-limiting audits.
\newblock {\em {IEEE} Security and Privacy}, 10:42--49, 2012.

\bibitem{LindemanStYa12}
M.~Lindeman, P.~B. Stark, and V.~S. Yates.
\newblock {BRAVO}: Ballot-polling risk-limiting audits to verify outcomes.
\newblock In {\em Proc. 2012 EVT/WOTE}, 2012.

\bibitem{NBHC07}
L.~Norden, A.~Burstein, J.~L. Hall, and M.~Chen.
\newblock Post-election audits: Restoring trust in elections.
\newblock Technical report, Brennan Center for Justice and Samuelson Law,
  Technology \& Public Policy Clinic, 2007.

\end{thebibliography}

\end{document}